\documentclass[11pt,preprint,aps,prc,showpacs,tightenlines]{revtex4}
\usepackage{epsf}
\usepackage{amsmath}

\def\beq{\begin{equation}}
\def\eeq{\end{equation}}
\def\bea{\begin{eqnarray}}
\def\eea{\end{eqnarray}} 
\def\beqa{\begin{equation}\begin{array}{l}}
\def\eeqa{\end{array}\end{equation}}
\def\eqlab#1{\label{eq:#1}}
\def\figlab#1{\label{fig:#1}}
\def\tablab#1{\label{tab:#1}}
   
\def\eref#1{(\ref{eq:#1})}
\def\Eqref#1{Eq.~(\ref{eq:#1})}

\def\Figref#1{Fig.~\ref{fig:#1}}
\def\tabref#1{\ref{tab:#1}}

\def\half{\mbox{\small{$\frac{1}{2}$}}}

\def\third{\mbox{\small{$\frac{1}{3}$}}}

\def\barr{\left(\begin{array}{c}}
\def\earr{\end{array}\right)}
\def\bmat{\left(\begin{array}{cc}}
\def\emat{\end{array}\right)}
\def\al{\alpha}
\def\be{\beta}
\def\ga{\gamma} 
 \def\De{\Delta}
\def\veps{\varepsilon}  

 \def\La{{\Lambda}}

\def\w{\omega}

\def\pa{\partial}

\def\Z{{\mathcal Z}}

\def\pa{\partial}

\def\nn{\nonumber}

\def\lag{{\mathcal L}}

\def\mathscr{\mathcal}

\def\3d{3-D}

\def\ol#1{\overline{#1}} 

\def\khat{{\bf q}}

\def\half{{\textstyle \frac 1 2}}

\def\epsbol{\mbox{\boldmath $\epsilon$}}
\def\sigbol{\mbox{\boldmath $\sigma$}}
\def\mn{M}

\begin{document}


\title{Model-independent effects of $\Delta$ excitation in nucleon
polarizabilities}
\author{Vladimir Pascalutsa}
\email{vlad@phy.ohiou.edu}
\author{Daniel R. Phillips}
\email{phillips@phy.ohiou.edu}

\affiliation{
Institute for Nuclear and Particle Physics (INPP),
Department of Physics and Astronomy, Ohio University,
Athens, OH 45701}
\date{\today}

\begin{abstract}
Model-independent effects of $\Delta$(1232) excitation on nucleon
polarizabilities are computed in a Lorentz-invariant fashion.  We find
a large effect of relative order $(M_\Delta - M)/M$ in some of the
spin polarizabilities, with the backward spin polarizability receiving
the largest contribution. Similar subleading effects are found to be
important in the fourth-order spin-independent polarizabilities
$\alpha_{E\nu}$, $\al_{E2}$, $\be_{M\nu}$, and $\be_{M 2}$.
Combining our results with those for the model-independent effects of
pion loops we obtain predictions for spin and fourth-order
polarizabilities which compare favorably with the results of a recent
dispersion-relation analysis of data. 
\end{abstract}

\maketitle
\thispagestyle{empty}


\section{Introduction}
The nucleon Compton scattering ($\gamma$N$\rightarrow \gamma$N) amplitude,
in Coulomb gauge,  can be
written in terms of six operator structures with coefficient
functions $A_1(s,t)$ to  $A_6(s,t)$:
\begin{eqnarray}
T_{fi}&=&\epsbol'\cdot\epsbol\,A_1(s,t) 
+\epsbol'\cdot\khat\,\epsbol\cdot\khat'\,A_2(s,t) \nonumber\\
&&+i\sigbol\cdot(\epsbol'\times\epsbol)\,A_3(s,t)+
i\sigbol\cdot(\khat'\times \khat)\,\epsbol'\cdot\epsbol \,
             A_4(s,t)\nonumber\\
&&+\Bigl(i\sigbol\cdot(\epsbol'\times \khat)\,\epsbol\cdot\khat'-
i\sigbol\cdot(\epsbol\times \khat')\,\epsbol'\cdot\khat\Bigr)\,
             A_5(s,t)\eqlab{amp}\\
&&+\Bigl(i\sigbol\cdot(\epsbol'\times \khat')\,\epsbol\cdot\khat'-
i\sigbol\cdot(\epsbol\times \khat)\,\epsbol'\cdot\khat\Bigr)\,
             A_6(s,t),\nonumber
\end{eqnarray}
where $\khat$ ($\khat'$) is the 
initial (final) three-momentum of the photon, and 
 $\epsbol$ ($\epsbol'$) is the photon polarization vector.
 
{\em Nucleon polarizabilities} can be defined as coefficients in the
low-energy expansion of this amplitude. Namely, expanding the
amplitude in powers of the photon energy $\omega$, the first two terms
[$O(\omega^0)$ and $O(\omega^1$)] are fixed by the low-energy theorem
(LET)~\cite{LET}, while terms of $O(\omega^2)$ are proportional to the
electric and magnetic polarizabilities $\alpha$ and $\beta$, terms of
$O(\omega^3)$ define the spin polarizabilities~\cite{Rag93} $\gamma_1$
to $\gamma_4$, and terms of $O(\omega^4)$ define the fourth-order
spin-independent polarizabilities $\al_{E\nu}$, $\be_{M\nu}$,
$\al_{E2}$, $\be_{M2}$~\cite{BGLMN,Hol00}.  More specifically, the
Breit-frame Compton amplitude---up to and including terms of order
$\w^4$---can be written using \Eqref{amp} and~\footnote{In
\Eqref{ampexp}, $M$ is the nucleon mass, ${\cal Z}=1$ for the proton,
${\cal Z}=0$ for the neutron, $e^2/4\pi\simeq 1/137$, and
$\kappa=(\kappa_s+\kappa_v\tau_3)/2$, with $\kappa_s\simeq -0.12$ and
$\kappa_v\simeq 3.71$ the isoscalar and isovector anomalous magnetic
moments of the nucleon.}:
\begin{eqnarray}
A_1(s,t)\!&=&\!-{{\cal Z}^2e^2\over\mn}
+ 4\pi\w\w'(\alpha+\be z)\nonumber\\
&& + 4\pi(\w\w')^2
\left[\alpha_{E\nu}+(\mbox{$\frac{1}{12}$}\al_{E2}+\be_{M\nu})\, z
+ \frac{1}{12}\be_{M2}(2z^2-1) \right] +
O(\omega^6)\nonumber\\
A_2(s,t)\!&=&\!-4\pi\beta - 4\pi\w\w' 
\left( \be_{M\nu} - \frac{1}{12}\al_{E2} + \frac{z}{6}\be_{M2}\right)
+ O(\omega^4)\nonumber\\
A_3(s,t)\!&=&\! \half (\w+\w')\left\{
{e^2 \over 2\mn^2}
\left[{\cal Z}({\cal Z}+2\kappa)-({\cal Z}+\kappa)^2 z\right] +
4\pi\omega\w'(\gamma_1+\gamma_5 z)\right\}+{O}(\omega^5)\nonumber\\
A_4(s,t)\!&=&\! \frac{\w+\w'}{2\w\w'}\left[
-{e^2\over 2\mn^2 }({\cal Z}+\kappa)^2 
+4\pi\w\w' \gamma_2\right] +{O}(\omega^3)\eqlab{ampexp}\\
A_5(s,t)\!&=&\! \frac{\w+\w'}{2\w\w'}\left[
{e^2 \over 2\mn^2 }({\cal Z}+\kappa)^2 
+4\pi\w\w'\gamma_4\right] +O(\omega^3)\nonumber\\
A_6(s,t)\!&=&\! \frac{\w+\w'}{2\w\w'}\left[
-{e^2 \over 2\mn^2 }{\cal Z}({\cal Z}+\kappa)
+4\pi\omega\w'\gamma_3\right] +O(\omega^3),\nonumber
\end{eqnarray}
where\footnote{ The quantities $\omega$, $\omega'$, and
$z$ are defined in an invariant fashion. In the lab frame they can be
interpreted as the initial and final photon energy and the scattering
angle.} $\w=(s-M^2)/2M$, $\w'=(M^2-u)/2M$, and
$z=(1+t/2\w\w')$. (Note that nucleon pole terms proportional to $\omega^2/M^3$
in $A_1$ and to $1/M^3$ in $A_2$ are not essential to what follows and
so have been omitted here.)

In other words, after the Compton amplitude is expanded in powers of
photon energy:
\beq
T_{fi} =\sum_k  T^{(k)} \w^k ,
\label{eq:wexp}
\eeq 
where the coefficients $T^{(k)}$ are operator valued,  \Eqref{amp}
can be presented schematically as:
\beq
T^{(0)} \sim \frac{(e{\cal Z})^2}{M}\,, \,\,\,
T^{(1)} \sim \frac{\kappa}{M^2} \,, \,\,\,
T^{(2)} \sim (\al,\be)\,, \,\,\,
T^{(3)} \sim (\ga_1,\ga_2,\ga_3,\ga_4),\,\,\,
T^{(4)} \sim (\alpha_{E\nu},\be_{M\nu},\al_{E2},\be_{M2})\,.
\eeq 
According to the LET, the Born graphs give the full result for
$T^{(0)}$ and $T^{(1)}$.  All other effects (e.g., meson loops)
 which contribute to $T^{(0)}$ and
$T^{(1)}$ can only result in renormalizations of the charge and
magnetic moment.

In contrast, $T^{(2)}$, $T^{(3)}$, etc., can be influenced by a number
of effects. Most significant are those that are proportional to
a negative power of light hadronic scales, the lightest such
scale being, of course,  the pion mass, $m_\pi\simeq 139$ MeV. Contributions which scale with
negative powers of $m_\pi$ are due to the long-range effects of the
pion cloud, see \Figref{pinloops}. These contributions to
$\alpha$, $\beta$, and the $\gamma_i$'s are known from chiral
perturbation theory~\cite{BKM,Manch00,Gel00}. 
\begin{figure}[t]
\vspace*{0.5cm}
\centerline{  \epsfxsize=13.5 cm
  \epsffile{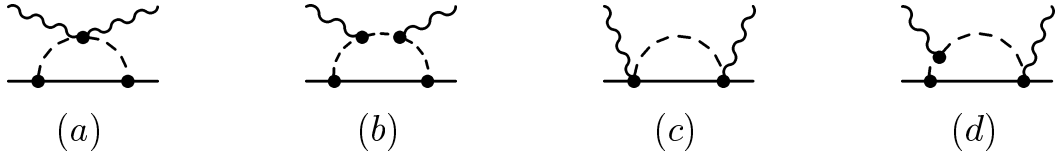}
}
\caption{The one $\pi$N-loop contributions to $\gamma$N 
scattering (crossed graphs are not shown).}
\figlab{pinloops}
\end{figure} 

The next lightest hadronic scale is the excitation energy of the
$\Delta(1232)$-isobar: \beq \eqlab{Delta} \De = M_\De -M\simeq
293\,\mbox{ MeV}.  \eeq The most significant effects of $\Delta$ excitation
contribute inverse powers of $\De$ to the nucleon polarizabilities,
and hence, diverge in the large-$N_c$ limit. In the real world where
$\De$ is not vanishing but given by \Eqref{Delta}, these effects are
still potentially important, since $\De$ is heavier than $m_\pi$ by
only about a factor of two.

The leading effect of $\Delta$ excitation was computed  for
$\alpha$, $\beta$, and $\gamma_1$--$\gamma_4$ by Hemmert {\it
et al.}~\cite{He97,He98B}, and for the fourth-order polarizabilities by
Holstein {\it et al.}~\cite{Hol00}. 
Here we shall compute the complete
effect of the $\Delta$-excitation contribution (\Figref{Delta}) in a manifestly
covariant fashion. While we find agreement with previous calculations
for the leading contributions, the {\it subleading} ones,  
suppressed by $\Delta/M$ relative to leading, 
bring a sizable correction to some polarizabilities.  The
biggest correction is in the {\it backward} spin polarizability,
$\ga_\pi$. In fact, there the subleading $\De$ effect exceeds the
leading one by {\em at least} a factor of two. We stress that this
result is a model-independent consequence of Lorentz invariance and
the existence of a light P33-resonance.  Indeed, any model of
Compton scattering should give the same answer for all
pieces of polarizabilities which scale with negative powers of
$m_\pi$ and $\Delta$. 
\begin{figure}[b]
\centerline{  \epsfxsize=10 cm
  \epsffile{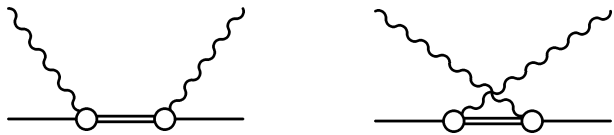}
}
\caption{The $\Delta$-excitation graphs.}
\figlab{Delta}
\end{figure} 

The paper is organized as follows. In Sec.~\ref{NDA} we perform naive
dimensional analysis on the operators $T^{(k)}$ of
Eq.~(\ref{eq:wexp}), in order to estimate the relative importance of
$\pi$N loops, $\De$ excitation, $\pi\De $ loops, and other mechanisms
in these quantities. Having established that polarizabilities receive
contributions which scale with negative powers of $m_\pi$ and $\De$,
we compute the $1/\De$ pieces arising from $\De$ excitation in
Sec.~\ref{De}. In Sec.~\ref{pi} we briefly review the results of
Refs.~\cite{BKM,Gel00,Manch00,He97,He98B,Hol00} for the pieces of
$\alpha$, $\beta$, $\gamma_1$--$\gamma_4$, and $\al_{E \nu}$,
$\al_{E2}$, $\be_{M\nu}$, $\be_{M2}$, which are due to $\pi$N and
$\pi\De$ loops. We sum the model-independent contributions discussed
in Secs.~\ref{De} and \ref{pi} and compare to results of a
dispersion-relation analysis as well as to recent experimental values for
the forward and backward spin polarizabilities in Sec.~\ref{conc}.

\section{Naive dimensional analysis for polarizabilities}
\label{NDA}

Contributions to $T^{(k)}$ in \Eqref{wexp} can be classified according
to whether they are generated by pion physics, by the excitation of
$\De$ degrees of freedom, or by physics at a higher-energy scale,
$\La$. The minimum value of $\La$ is set by the mass of the next light
meson or by the next $N^\ast$-resonance excitation-energy. $\La$ can also
take values of the other heavy-mass scales in the theory, such as $M$,
$M_\De$, and $4\pi f_\pi$.

At tree level pions can only contribute to Compton scattering through
the chiral Wess-Zumino-Witten (WZW) anomaly.  
The leading (in negative powers of $m_\pi$)
contribution of the WZW-anomaly  to the
amplitude $T^{(k)}$ ($k\ge 3$) scales as:
\begin{subequations}
\eqlab{pc}
\beq
\eqlab{pc0}
T^{(k)}(\mbox{Anomaly}) \sim 
\frac{1}{\La^{2} m_\pi^{k-1}}\,.
\eeq
In the case of loop graphs the dominant contribution of a pion-nucleon
$L$-loop graph to the amplitude $T^{(k)}$, with $k\ge L+1 $, behaves
according to:
\beq
\eqlab{pc1}
T^{(k)}(\mbox{$\pi N$ loop}) \sim 
\frac{1}{\La^{2L} m_\pi^{k+1-2L}}\,.
\eeq
Meanwhile, the leading contribution to $T^{(k)}$
due to $\Delta$ excitation, i.e. the one with most powers of $1/\De$, is
a tree-level graph with the $\De$ in either the $s$- or $u$-channel (\Figref{Delta}):
\beq
\eqlab{pc2}
T^{(k)}(\De) \sim 
\frac{1}{\La^{2} \De^{k-1}}\,,
\eeq 
as long as $k \geq 2$. The 
 contribution of a $\pi\De$ $L$-loop graph has as its leading piece:
\beq
\eqlab{pc3}
T^{(k)}(\mbox{$\pi \De$ loop}) \sim \left\{\begin{array}{cc}
\frac{1}{\La^{2L} m_\pi^{k-2L}\De} \,, & \mbox{even $k$;}\\
\frac{1}{\La^{2L} m_\pi^{k-2L-1} \De^{2}} \,, & \mbox{odd $k$;}
\end{array} \right.
\eeq
provided that $k\ge L+1 $ and we assume $m_\pi$ is 
significantly smaller than $\De$.
\begin{figure}[t]
\vspace*{0.5cm}

\centerline{  \epsfxsize=13.5 cm
  \epsffile{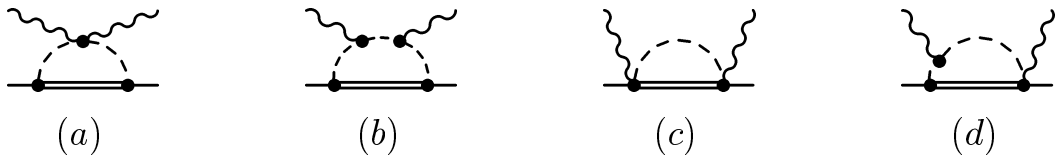}
}

\caption{The one-$\pi \Delta$-loop contributions to nucleon polarizabilities.}
\figlab{deltaloop}
\end{figure} 

Finally, all the higher energy (``short-range'') effects  scale as:
\beq
\label{pc4}
 T^{(k)} (\mbox{short-range})\sim \frac{1}{\La^{k+1}}\,.
\eeq
\end{subequations}
Therefore, if $\La$ is significantly above $m_\pi$ and $\De$,
the short-range physics cannot affect
the contributions which scale with negative powers of $m_\pi$ and $\De$.
Hence the latter contributions are not only dominant at low energy
but also model-independent. 
Any theory of Compton scattering which obeys chiral, gauge, and Lorentz
symmetries, includes
pion loops, and has a light $\De$-resonance
should give the same answer for the contributions which
scale with negative powers of $m_\pi$ and $\De$.


\section{Calculation of $\De$-excitation effects}

\label{De}
To compute the effect due to $\De$ excitation,
we assume the following form of the electromagnetic $N\De$ transition 
Lagrangian~\cite{PP}:
\beq
\eqlab{ganDe}
\lag_{\ga N \De}= \frac{3\,e}{4 M M_+}\,\ol N\, T_3^\dagger
\left(i g_M  \tilde F^{\mu\nu}
- g_E \gamma_5 F^{\mu\nu}\right)\,\pa_{\mu}\De_\nu
+ \mbox{H.c.},
\eeq
where $M_+=\half (M+M_\De)$, and $T_3$ is the isospin $N\De$ transition
matrix, with normalization $T^\dagger_3 T_3=\frac{2}{3}$.

This $\ga N\De$ coupling is  invariant under electromagnetic gauge 
transformations (to the order to which we work), as well as
under  the spin-3/2 gauge transformation:
\begin{equation}
\De_\mu(x) \rightarrow \De_\mu(x) + \pa_\mu \veps(x),
\label{eq:spin3/2gt}
\end{equation}
with $\veps$ a spinor field. Invariance under (\ref{eq:spin3/2gt}) ensures
the correct spin-degrees-of-freedom counting~\cite{Pa98}.  Other
forms of this coupling, such as the conventional $G_1$-$G_2$
representation with off-shell parameters~\cite{PaS95}, may result in
different ``short-range'' pieces of polarizabilities, however 
contributions proportional to negative powers of $\Delta$ will be the same.

In the Delta's rest frame  (where $\De_0=0$, $ \pa_0 \De_i=-i M_\De \De_i$,  
and $\pa_i \De_j=0$) the coupling \eref{ganDe} becomes
\beq
\eqlab{ganDe2}
\lag_{\ga N \De}=  - \frac{3\,e M_\De}{4 M M_+}\,\ol N\, T_3^\dagger
\left( g_M  B^i
+ g_E \gamma_5 E^i \right)\, \De_i
+ \mbox{H.c.},
\eeq
where $B^i$ is the magnetic and $E^i$ the electric field.
Thus, the two terms correspond to  $N\De$ magnetic
and electric transitions, respectively. The precise relation of these
couplings to the conventional helicity and multipole
amplitudes is given in the Appendix.

Computing the sum of the $s$- and $u$-channel  $\Delta$
contributions, \Figref{Delta}, to the polarizabilities we obtain
(see Ref.~\cite{PP} for more details):
\begin{subequations}
\eqlab{result}
\bea
(\al,\, \be) &=&  \frac{e^2}{4\pi} \frac{1}{2M_+^2}\, 
\left(-\frac{g_E^2}{2M_+}  ,\,\, \frac{g_M^2}{\De}  \right) \,,\\
(\ga_1,\, \ga_2, \,\ga_3,\, \ga_4)
 &=& 
\frac{e^2}{4\pi} \frac{M}{4 M_+^3 \De}
\left(\frac{4M_+}{M^2}
g_M^2 + \frac{g_E^2}{2 M_+} ,\,\,  -\frac{g_M^2}{\De},\,\,-
\frac{2M_+}{M^2}g_M^2,\,\,
\frac{M_\De^2}{M^2}\frac{g_M^2}{\De}\right). 
\eea
\end{subequations}

This is an exact result for the $\De$-excitation contribution.  The result
of Hemmert {\it et al.}~\cite{He97,He98B} corresponds to the leading
term in a $\De/M$ expansion of these expressions.  We choose to expand
in $\De/M_+$, then $\be$ is entirely of leading order (LO) while $\al$ is of
next-to-leading order (NLO):\footnote{Note that in general, whether the leading contribution
is due to a magnetic or electric transition will be determined by the parity
of the resonance.}
\begin{subequations}
\bea
\al &=&-\frac{e^2}{4\pi} \frac{g_E^2}{4M_+^3}=O(1),\\
\be&=&\frac{e^2}{4\pi} \frac{g_M^2}{2M_+^2\,\De}=O(1/\De)\,.
\eea
\end{subequations}
The spin polarizabilities to, respectively, LO and NLO in this expansion
are given by:
\begin{subequations}
\bea
O(1/\De^2) &:\,&(\ga_1,\, \ga_2, \,\ga_3,\, \ga_4)
=\frac{e^2}{4\pi} \frac{M}{4M_+^3\De^2} 
\left(0 ,\,  -g_M^2,\,0,\,g_M^2\right). \eqlab{delga1}\\
O(1/\De) &:\,&(\ga_1,\, \ga_2, \,\ga_3,\, \ga_4)
 =\frac{e^2}{4\pi} \frac{1}{4 M_+^2 M \De} 
\left(4 g_M^2 +\frac{M^2}{2M_+^2}g_E^2,\,\,  0,\,\,-2 g_M^2,\,\,
2 g_M^2\right). \eqlab{delga2}
\eea
\end{subequations}

Of special interest are the forward ($\ga_0\equiv 
\ga_1 -\ga_2-2\ga_4 $) and backward ($\ga_\pi\equiv \ga_1 +\ga_2+2\ga_4$) spin
polarizabilities. From \Eqref{delga1} and \Eqref{delga2} it is easy to derive
the Delta contribution to these quantities: 
\begin{subequations}
\bea
O(1/\De^2) &:\,& 
(\ga_0,\, \ga_\pi)  = 
\frac{e^2}{4\pi}\frac{M}{4 M_+^3 \De^2} ( -g_M^2,\,g_M^2),\\
O(1/\De) &:\,& 
(\ga_0,\, \ga_\pi)  = 
\frac{e^2}{4\pi}\frac{M}{4 M_+^4 \De} \left( -\half g_E^2,
\, \frac{8M_+^2}{M^2} g_M^2 + \half g_E^2\right).
\label{eq:domde}
\eea
\end{subequations}

Our expansion parameter $\De/M_+\sim \third$ is relatively
small.  However, the subleading result for the spin polarizabilities
contains large coefficients. In fact, in the backward direction, the
subleading effect is always larger than the leading one by at least a
factor of \beq \frac{\ga_\pi^{\mathrm{NLO}}}{\ga_\pi^{\mathrm{LO}}} =
8 \frac{\De M_+}{M^2}\approx 2\,.  \eeq Non-zero values of $g_E$ 
only serve to increase this factor.

Next we compute the $\De$-excitation contribution to the
fourth-order polarizabilities at leading and
subleading order. Details are given in Appendix B. The result is:
\begin{subequations}
\bea
O(1/\De^3) &:\,&(\alpha_{E\nu},\be_{M\nu},\al_{E2},\be_{M2})
=\frac{e^2}{4\pi} \frac{1}{M_+^2\De^3} 
\left(0 ,\,  g_M^2,\,0,\,0\right), \eqlab{p4th1}\\
O(1/\De^2) &:\,&(\alpha_{E\nu},\be_{M\nu},\al_{E2},\be_{M2})
 =\frac{e^2}{4\pi} \frac{1}{4 M_+^3 \De^2} \nn\\
&&\times\left(-\mbox{$\frac{13}{2}$} g_M^2 -g_E^2 +2g_E g_M,\,  
g_M(g_M-g_E),\,0,\,-6 g_M^2\right). \eqlab{p4th2}
\eea
\end{subequations}

\section{Review of model-independent pion-loop contributions}
\label{pi}

All the $1/m_\pi$ terms which scale as in \Eqref{pc0} and \Eqref{pc1}
have already been computed~\cite{BKM,Manch00,Gel00}.
The leading non-analytic (LNA) behavior in $m_\pi$ of $\alpha$ and $\beta$
comes entirely from one-loop graphs, \Figref{pinloops}, in chiral perturbation 
theory~\cite{BKM}:
\beq
O(1/m_\pi) \,:\,
(\al,\, \be) = \frac{e^2}{4\pi}\left(\frac{g_A}{4\pi f_\pi}\right)^2\frac{1}{m_\pi}  
\,\left(\frac{5\pi}{6}, \, \frac{\pi}{12}\right),
\eeq
where $e^2/4\pi \simeq 1/137$, $g_A\simeq 1.26$,
$f_\pi\simeq 93$ MeV. Corrections to $\alpha$ and $\beta$ suppressed
by $m_\pi/M$ relative to leading have the same scaling as short-range
effects (\ref{pc4}), and so a model-independent result for them
is less useful.

For the fourth-order spin polarizabilities the situation is
different, since the leading contribution has an additional
power of $1/m_\pi$~\cite{BKM}:
\bea 
O(1/m_\pi^2) &:& (\ga_1,\,\ga_2,\,\ga_3,\,\ga_4) =
\frac{e^2}{4\pi}\left[\frac{1}{3m_\pi^2}\left(\frac{g_A}{4\pi f_\pi}\right)^2
\,\left(2, \, 1,\, \frac{1}{2},\, -\frac{1}{2}\right) \right. \nn\\ 
&& \hskip3cm + \left. \frac{g_A(2\Z-1)}{(2\pi f_\pi)^2 m_{\pi^0}^2}
\left(-1,0,\frac{1}{2},-\frac{1}{2}\right) \right]
\,.  \eea
In this case $\pi$N loops and the WZW anomaly graph
all contribute.

Corrections of $O(m_\pi/M)$ relative to these leading effects are then
also model-independent predictions of chiral perturbation
theory. These were computed recently in~\cite{Manch00}:
\begin{subequations}
\bea
O(1/m_\pi) \,& : & (\ga_1,\,\ga_2,\,\ga_3,\,\ga_4) =
-\frac{e^2}{4\pi}\left(\frac{g_A}{4\pi f_\pi}\right)^2
\frac{\pi}{12 M m_\pi}
 \nn\\
&& \times \left( 3 + 10{\cal Z}, \, 8+ \kappa_v + 3(1+\kappa_s)(2{\cal Z} -1),\,  
\right. \label{eq:manch}\\
&& \left. \,\,\,\mbox{$\frac{5}{2}$}+{\cal Z},\,
 -\mbox{$\frac{15}{2}$}-2\kappa_v - 2(1+\kappa_s)(2{\cal Z}-1)\,\right) 
\nn
\eea
and in~\cite{Gel00}:
\bea
O(1/m_\pi) &:& (\ga_1,\,\ga_2,\,\ga_3,\,\ga_4) =
-\frac{e^2}{4\pi}\left(\frac{g_A}{4\pi f_\pi}\right)^2
\frac{\pi}{12 M m_\pi}\nn\\
&&\times \left( 3 + 10{\cal Z}, \, 6- \kappa_v +(1+\kappa_s)(2{\cal Z} -1),\,  
\mbox{$\frac{5}{2}$}-{\cal Z},\,
 - \mbox{$\frac{11}{2}$}\right).
\label{eq:julich}
\eea
\end{subequations}
Although the results of \cite{Gel00} and \cite{Manch00} are different,
the computation of all one-loop graphs in both papers agree, as
do the predictions for all directly-observable experimental
quantities~\cite{comment}.  The difference lies in the definition of spin
polarizabilities.  

For the fourth-order polarizabilities only the LNA contribution
of $\pi N $ loops is known at present~\cite{BKM,BGLMN,Hol00}:
 \beq 
O(1/m_\pi^3) :\, (\alpha_{E\nu},\be_{M\nu},\al_{E2},\be_{M2}) =
\frac{e^2}{4\pi}\left(\frac{g_A}{4\pi f_\pi}\right)^2\frac{\pi}{10\,m_\pi^3}
(\mbox{$\frac{3}{4}$},\,\,\mbox{$\frac{7}{6}$},\,\,
7,\,\,-3)\,.
\eeq

Contributions that are proportional to negative powers of $\De$ can
also come from the $\pi\De$-loop graphs, \Figref{deltaloop}.  The
scaling (\ref{eq:pc3}) means that $\pi \De$ loops contribute to the
$O(1/\De)$ term in $\alpha$ and $\beta$, and to both the $1/\De^2$ and
$1/\De$ term in the spin polarizabilities. The contribution to the
spin-independent polarizabilities was computed in Ref.~\cite{He97}:
 \beq
O(1/\De) : (\al,\, \be) =  \frac{e^2}{4\pi} 
 \frac{2}{3}\left(\frac{h_A}{4\pi f_\pi}\right)^2
\,\frac{1}{\De}\left(1+\mbox{$\frac{1}{9}$}\ln f(\mbox{$\frac{\De}{m_\pi}$}),\,  
\mbox{$\frac{1}{9}$}\ln f(\mbox{$\frac{\De}{m_\pi}$})\right)\,,
\eeq
with $f(\xi) = \xi + \sqrt{\xi^2-1}$ and $h_A$ the $\pi N\De$ coupling. 
The LNA behavior of the $\pi \De$
loops for the $\gamma$'s has been computed in
Ref.~\cite{He98B}. It is:
\bea
O(1/\De^2 ) \,: (\ga_1,\, \ga_2, \,\ga_3,\, \ga_4)
&=&\frac{e^2}{4\pi} 
\frac{1}{27\De^2}\left(\frac{h_A}{4\pi f_\pi}\right)^2 \nn\\
&\times & \left(-2,\,
 2- 2\ln f(\mbox{$\frac{\De}{m_\pi}$}),\, 1- \ln f(\mbox{$\frac{\De}{m_\pi}$})\,,
-1+ \ln f(\mbox{$\frac{\De}{m_\pi}$})\right).
\label{eq:pideloops}
\eea
Unfortunately, the $1/\De$ piece of the $\pi \Delta$ loop effect on
$\gamma_1$--$\gamma_4$ does not yet exist in the
literature. Since the leading contribution of $\pi \De$ loops 
(\ref{eq:pideloops}) is numerically
small, we expect the $1/\De$ piece to be small as well.  Nevertheless, a future
computation of this contribution is important as it will complete the analysis of the
model-independent effects in spin polarizabilities.

For the fourth-order polarizabilities situation is even less satisfactory.
There we know completely just the $O(1/m_\pi^2\De)$ and $O(1/\De^3)$
contributions due to $\pi \De$ loops~\cite{Hol00}:
\begin{subequations}
\bea
O(1/m_\pi^2\De) &:& (\alpha_{E\nu},\be_{M\nu},\al_{E2},\be_{M2})
= \frac{e^2}{4\pi} 
\frac{2}{45 m_\pi^2\De}\left(\frac{h_A}{4\pi f_\pi}\right)^2 
\left( -\mbox{$\frac{29}{18}$},\, \mbox{$\frac{3}{2}$}, \, \mbox{$\frac{22}{3}$},\,-6 \right), \\
O(1/\De^3) &:& (\alpha_{E\nu},\be_{M\nu},\al_{E2},\be_{M2})
= \frac{e^2}{4\pi} 
\frac{2}{45 \De^3}\left(\frac{h_A}{4\pi f_\pi}\right)^2 
\left(  \mbox{$\frac{56}{18}$} - \mbox{$\frac{1}{6}$}\ln f(\mbox{$\frac{\De}{m_\pi}$}), \right. \nn\\
&& \left. \, -1+ \mbox{$\frac{1}{6}$}\ln f(\mbox{$\frac{\De}{m_\pi}$}),\,
-12+ 6 \ln f(\mbox{$\frac{\De}{m_\pi}$}), \, 6- 6\ln f(\mbox{$\frac{\De}{m_\pi}$}) \right). 
\eea
\end{subequations}
One-loop graphs with insertions from ${\cal L}_{\pi \De N}^{(2)}$ can
generate effects in the fourth-order polarizabilities scaling like
$1/(m_\pi \De M_+)$. Results for these sub-leading effects do not
presently exist in the literature. However, we expect them to be
comparable to the small $1/\De^3$ effects calculated in
Ref.~\cite{Hol00}, since $\De/M_+$ is of roughly the same size as
$m_\pi^2/\De^2$. Once again though, checking this expectation in a
full calculation of these ``relativistic'' $\pi \De$-loop effects is
an important future step.

\section{Discussion and conclusion}
\label{conc}

Combining the results of previous two sections, we now have all the
$1/m_\pi$ and $1/\De$ pieces of nucleon spin polarizabilities, except one
--- the subleading [i.e., $O(1/\De)$] contribution of the $\pi\De$
loops. This is expected to be small, so here
we focus on the model-independent contributions which are already
known.

The numerical values for these pieces of $\gamma_1$--$\gamma_4$ are
presented in Table~\tabref{pols0}.  The sum of these contributions can
be compared to the results of the dispersion-relation (DR) analyses
shown in the last two columns. The differences between the two DR
analyses can be regarded as indicative of the size of their
theoretical uncertainty. For the $\pi N$-loop $O(m_\pi^{-1})$
contribution we quote two results: the first is due to
Ref.~\cite{Gel00} and the second (in brackets) is due to
Ref.~\cite{Manch00}.  The total sum also is given as two numbers in
the cases where \cite{Gel00} and \cite{Manch00} disagree.

In generating Table~\tabref{pols0} for the $\ga$N$\De$ couplings we
used the values extracted from our recent analysis of Compton
scattering data~\cite{PP}: $g_M =2.6 $, $g_E=-6$.  The value of $g_M$
is consistent with the large-$N_c$ value $g_M\simeq 2.63$.  
The value of $g_E$ is unusually
large, however,  combined with the rather small $g_M$ value, 
it leads to a reasonable radiative width of the $\De$ resonance.
Also, here $g_E$ affects only $\ga_1$ at $O(\De^{-1})$---and that in
a fairly mild way: the magnetic transition still dominates over the 
electric one
in spin polarizabilities.  For the $\pi N\De$
coupling we have used the large-$N_c$ estimate:
$h_A=\frac{3}{\sqrt{2}}g_A\simeq 2.7$, which is about $5\%$ smaller
than the value inferred from the width of the $\De$-resonance.

From this table it is clear that adopting the results of
Ref.~\cite{Gel00} for the $O(m_\pi^{-1})$ corrections makes the
comparison to the DR analysis much more favorable. It is
also clear that the $O(\De^{-1})$ effect plays a crucial role in
achieving  agreement with the DR result.

Table~\tabref{pols} shows the model-independent contributions for the
forward and backward spin polarizabilities of the nucleon.  
The sum of all the presented contributions can again be compared to
DR analyses  and also to the recent experimental
results obtained at the LEGS (BNL) and MAMI (Mainz) facilities.

The $O(1/\De)$ effect of the $\De$-excitation plays a very significant
role in the backward spin polarizability $\ga_\pi$. Because of this
large and positive correction the sum of all model-independent pieces
for the proton is $\ga_\pi^{(p)}=-34\times 10^{-4}$ fm$^4$, which lies in
between the mutually-inconsistent LEGS and MAMI measurements. The
prediction for the neutron is consistent with the recent MAMI
measurement.

In Table~\tabref{pols4} we show the results for some of the
model-independent contributions to the fourth-order
polarizabilities. Their sum can then be compared with two recent
dispersion analyses. While the leading $\De$-excitation effect is
non-vanishing only for $\be_{M\nu}$, the subleading contribution is
substantial in $\al_{E\nu}$, $\be_{M\nu}$, and $\be_{M2}$. In all
three cases it helps produce better agreement with the DR results.

Admittedly, although the expressions for the polarizabilities
presented here are model independent, the particular values of the
parameters used depend on the method of their extraction.  For
instance, assuming that $\De$ excitation completely dominates the $E2$
and $M1$ multipoles of pion photoproduction at the $\De$ peak, one can
use the Particle Data Group values for the helicity
amplitudes~\cite{PDG} and the relations given in Appendix A to find
$g_M \simeq 3$ and $g_E\simeq -1$.  The prediction for all the
polarizabilities then changes accordingly, as shown in
Table~\tabref{altpols} where we present results for the
polarizabilities when the PDG $\ga N \De$ parameters are adopted in
preference to those found in the fit of Ref.~\cite{PP}. The subleading
effect of $\Delta$ excitation are important for this set of parameters
too.

We close with a note of caution. It is only fair to point out that,
while the sum of the model-independent contributions to the spin
polarizabilities presented here compares favorably with dispersion
relations and experiments, the situation in the spin-independent $\al$
and $\be$ polarizabilities is not as pleasing. There the
$O(m_\pi^{-1})$ effect alone is in good agreement with
experiment. This agreement is spoiled when
$O(\De^{-1})$ corrections due to $\De$ excitation and $\pi\De$ loops
are included.

Regardless of whether the agreement shown in Tables~\tabref{pols0} and
\tabref{pols} is coincidental or not, Eqs.~\eref{delga1} and
\eref{delga2} derived here are model-independent results for the
pieces of the nucleon polarizabilities arising from magnetic and
electric excitation of the $\De$. They exhibit a large 
correction (of order $1/\De$) to the leading result (of order
$1/\De^2$) for the spin polarizabilities and also 
produce substantial effects of $O(1/\De^2)$ in the fourth-order
polarizabilities.

\begin{acknowledgments}
We thank Gast\~ao Krein and Barbara Pasquini for useful
conversations regarding the dispersion-relation results. 
This research was supported by the U.~S. Department of
Energy under grants DE-FG02-93ER40756, DE-FG02-02ER41218, and by the
National Science Foundation under grant NSF-SGER-0094668.
\end{acknowledgments}

\appendix

\section{Helicity and multipole $\ga N\rightarrow\De$ amplitudes}
Here we relate $g_E$ and $g_M$ to conventional 
$\ga N\De$  amplitudes. The relation to the
helicity amplitudes is:
\begin{subequations}
\bea
A_{1/2} &=& -\frac{e}{8 M^{3/2}}\sqrt{\frac{\De}{M_+}}\left[
2 M_+\, g_M  +  \De\,  g_E\right]\, , \\
A_{3/2} &=& -\frac{\sqrt{3}\,e}{8 M^{3/2}}\sqrt{\frac{\De}{M_+}}\left[
 2M_+\, g_M- \De\, g_E \right]\, ,
\eea
\end{subequations}
where $\De=M_\De-M$, $M_+=\half (M_\De+M)$.
The inverse relation is:
\begin{subequations}
\eqlab{invrel}
\bea
g_E &=& - \frac{1}{e}\left( \frac{2 M}{\De}\right)^{3/2}
\sqrt{2M_+} \left( A_{1/2} - \frac{1}{\sqrt{3}}A_{3/2}\right) \,,\\
g_M &=& - \frac{1}{e}\frac{(2M)^{3/2}}{\sqrt{2M_+\De}}
\left( A_{1/2} + \frac{1}{\sqrt{3}}A_{3/2}\right)\,.
\eea
\end{subequations}
The relation to the multipole amplitudes is: 
\begin{subequations}
\bea
E2&=& -\half \left( A_{1/2} - \frac{1}{\sqrt{3}}A_{3/2}\right) =
\frac{e}{8 M^{3/2}}\sqrt{\frac{\De}{M_+}} 
\,\De\, g_E \,,\\
M1&=& -\half \left(A_{1/2} + \sqrt{3} A_{3/2}\right)=
\frac{e}{8 M^{3/2}}\sqrt{\frac{\De}{M_+}}\left[
 4 M_+\, g_M- \De\, g_E \right]\,.
\eea
\end{subequations}
Therefore the E2/M1 ratio is given by 
\beq
\eqlab{REM}
\frac{E2}{M1}=\frac{\frac{\De}{4M_+}g_E}{g_M -  \frac{\De}{4M_+}g_E}\,.
\eeq
It is interesting to note here that the correct 
large $N_c$ limit of this ratio
is trivially  reproduced by \Eqref{REM} 
provided that $g_E$ and $g_M$ do not depend on $N_c$.
Indeed, then only the masses depend on $N_c$:
$\De$ = $O(1/N_c)$, $M_+ = O(N_c)$,  and
hence, from \Eqref{REM}, the ratio behaves as 
\beq E2/M1 = O(1/N_c^2),
\eeq
in agreement with the recent result of Jenkins, Ji and Manohar~\cite{JJM}.

\section{On calculation of fourth-order polarizabilities}
Since the representation of the Compton amplitude given by
Eqs.~\eref{amp} and \eref{ampexp} is manifestly invariant under crossing ($s\leftrightarrow u$),
we compute only the s-channel contributions $A_i^{s}(\w,t)$. The
full amplitudes are then obtained by adding the crossed partner
as follows:
\bea
A_i(\w, t)&=& A_i^{s}(\w,t)+A_i^{s}(-\w',t), \,\,\,  \mbox{for $i=1,2$;} \nn\\
A_i(\w, t)&=& A_i^{s}(\w,t)-A_i^{s}(-\w',t), \,\,\,  \mbox{for $i=3,\ldots,6$;} \nn
\eea
Thus, if the $s$-channel amplitude has a low-energy expansion,
\beq
A_i^{s}(\w,t) = \sum_{nl} a_{nl}^{(i)}\, \w^n\,t^l
\eeq
the full amplitude expands as
\bea
A_i(\w, t)&=& \sum_l\!\left(\sum_{{\rm even}\,\, n} a_{nl}^{(i)}\, F_n +
\sum_{{\rm odd}\,\, n} a_{nl}^{(i)}\, G_n \right) t^l,\,\,\,  
\mbox{for $i=1,2$}\nn\\
A_i(\w, t)&=& \sum_l\!\left(\sum_{{\rm odd}\,\, n} a_{nl}^{(i)}\, F_n +
\sum_{{\rm even}\,\, n} a_{nl}^{(i)}\, G_n \right) t^l,\,\,\,  
\mbox{for $i=3,\ldots,6$} \nn
\eea
where $F_n \equiv \w^n+{\w'}^n$ and  $G_n \equiv \w^n-{\w'}^n$
satisfy the following recursion relations
\begin{subequations}
\bea
F_n &=& \w\w'\, (F_{n-2} +\tau \, G_{n-1}), \\
G_n &=& \w\w'\, (G_{n-2} + \tau\, F_{n-1}), 
\eea
\end{subequations}
with $\tau =- t/2\w\w'=1-z$. 
The coefficients $a_{nl}$ can then be straightforwardly related to
the polarizabilities defined by \Eqref{ampexp}.
For example, in fourth order we find the following relations:
\bea
&& \al_{E\nu} + \be_{M\nu} + \mbox{$\frac{1}{12}$}(\al_{E2} + \be_{M2})
= 2 a^{(1)}_{40}\,,\nn\\
&& \be_{M\nu} + \mbox{$\frac{1}{12}$}\al_{E2}+\third \be_{M2}
= 4 a_{21}^{(1)} -3 a_{30}^{(1)}\,, \nn\\
&& \be_{M\nu} -  \mbox{$\frac{1}{12}$}\al_{E2}+\mbox{$\frac{1}{6}$} \be_{M2}
= - 2a_{20}^{(2)} \,,\\
&&\mbox{$\frac{1}{6}$} \be_{M2}= 8 a_{02}^{(1)} + a_{20}^{(1)} - 2 a_{11}^{(1)}
= a_{10}^{(2)}-4 a_{01}^{(2)}\,.\nn
\eea
Our results for $A_i^{s}(\w,t)$ due to $\De$ excitation can be
found in Ref.~\cite{PP}. The resulting expressions for the
fourth-order polarizabilities are given above, in Eqs.~\eref{p4th1}
and \eref{p4th2}.

\newpage
\begin{table}[htb]
\begin{tabular}{||c|r|r|c|r|r|r|c||r|r||}
\hline
& WZW & \multicolumn{2}{|c|}{$\pi N$ loops \cite{Gel00} (\cite{Manch00})}
& \multicolumn{2}{|c|}{$\De$ excitation} & $\pi \De$ loops
& Sum & \multicolumn{2}{|c||}{Disp.\ relation} \\
\cline{2-7} \cline{9-10}
 $\ga^{(N)}_i$  & $O(m_\pi^{-2})$
&  $O(m_\pi^{-2})$ &  $O(m_\pi^{-1})$ & $O(\De^{-2})$ & $O(\De^{-1})$
& $O(\De^{-2})$  & & Ref.~\cite{DKH98} & Ref.~\cite{DPV03}\\
\hline
$\ga_1^{(p)} $ & $-22.1$ & $4.4$ & $-3.4$ ($-3.4$)&  0 & $3.5$ & $-0.5$  
 & $-18.1$ &  $-17.4$ & $-17.8$\\
$\ga_2^{(p)} $ & 0  & $2.2$  & $-0.8\,$($-3.7$)& $-1.8$ & $0$ & $-0.2$  
 &$-0.6\,$($-2.5$)  & $-1.1$ &$-0.8$\\
$\ga_3^{(p)} $ & $11.0$  &  $1.1$ & $-0.4\,$($-0.9$) & $0$ & $-1.2$  
& $-0.1$   & $10.5\,$($10.0$) & $10.6$ & 11.0\\
$\ga_4^{(p)} $ & $-11.0$  & $-1.1$& $1.4\,$($4.3$)&  $1.8$ & $1.2$  
& $0.1$&  $-7.6\,$($-4.6$) & $-7.6$ & $-8.1$ \\\hline
$\ga_1^{(n)} $ & $22.1$ & $4.4$ & $-0.8$ ($-0.8$)&  0 & $3.5$ & $-0.5$  
  & 28.7 & $28.6$ & 28.9 \\
$\ga_2^{(n)} $ & 0 &$2.2$  & $-0.4\,$($-2.4$) &  $-1.8$ & $0$ & $-0.2$  
   &$-0.2\,$($-2.2$) &$-0.8$ & $-0.7$\\
$\ga_3^{(n)} $ & $-11.0$  &  $1.1$ & $-0.7$ ($-0.7$) & $0$ & $-1.2$  
& $-0.1$ &  $-11.8$ & $-11.8$ & $-11.9$ \\
$\ga_4^{(n)} $ & $11.0$  & $-1.1$&  $1.4\,$($3.4$) & $1.8$ & $1.2$  
& $0.1$  & $14.4\,$(16.4) & $14.6$ & 14.8 \\\hline\hline
\end{tabular}
\caption{Pieces of nucleon spin polarizabilities which scale with
negative powers of light hadronic scales $m_\pi$ and $\De$ compared to
the dispersion-relation analysis of Ref.~\cite{DKH98}. The numbers in
round brackets in columns ``$O(m_\pi^{-1})$'' and ``Sum'' are 
obtained using \Eqref{manch}, while the numbers outside the brackets
are obtained using \Eqref{julich}. All values are in units of
$10^{-4}$ fm$^4$.}  \tablab{pols0}
\end{table}

\begin{table}[htb]
\begin{tabular}{||c|r|r|r|r|r|r|r||r|r|r||}
\hline
& WZW & \multicolumn{2}{|c|}{$\pi N$ loops \cite{Gel00}}
& \multicolumn{2}{|c|}{$\De$ excitation} & $\pi \De$ loops
& Sum & DR \cite{DKH98}& \multicolumn{2}{|c||}{Experiment} \\
\cline{2-7} \cline{10-11}
$\ga^{(N)}_i$  & $O(m_\pi^{-2})$
&  $O(m_\pi^{-2})$ &  $O(m_\pi^{-1})$ & $O(\De^{-2})$ & $O(\De^{-1})$ 
& $O(\De^{-2})$  &  &
 &  LEGS~\cite{LEGS01} &MAMI~\cite{MAMI01}\\
\hline
$\ga_0^{(p)}$ & 0& $4.4$ & $-5.4$ &$-1.8$ &1.2& $ -0.5$
&$-2.1$ &$-1.1$ & $-1.55 \pm 0.18$ & $-1.0 \pm 0.2$ \\
$\ga_\pi^{(p)}$ & $-44.1$ &$4.4$
& $-1.3$ &  $1.8$ & 5.8 &  $-0.5$ & $-34.0$ &$-33.7$ &$-27.2\pm 3.1$ &$-38.7\pm 1.8$\\
$\ga_0^{(n)}$ & 0& $4.4$ & $-3.3$ &$-1.8$ &1.2& $-0.5$& 0.0 &$0.2$ & &\\
$\ga_\pi^{(n)}$ & $44.1$ &$4.4$ &1.7
&  $1.8$ & 5.8 &  $-0.5$& 57.4 &57.0 & &$58.6\pm 4.0$\\
\hline\hline
\end{tabular}
\caption{Pieces of nucleon forward and backward 
spin polarizabilities which scale with negative 
powers of $m_\pi$ and $\De$ compared to a
dispersion-relation analysis and experimental values. 
(For the $O(m_\pi^{-1})$ contribution the
result of Ref.~\cite{Gel00} was used here.)
All values are in units of $10^{-4}$ fm$^4$.}
\tablab{pols}
\end{table}

\begin{table}[htb]
\begin{tabular}{||c|r|r|r|r|r|r||r|r||}
\hline
& $\pi N$ loops
& \multicolumn{2}{|c|}{$\De$ excitation} &\multicolumn{2}{|c|}{ $\pi\De$ loops}
&Sum & \multicolumn{2}{|c||}{Disp.\ relation} \\
\cline{2-6} \cline{8-9}
& $O(m_\pi^{-3})$ & $O(\De^{-3})$ & $O(\De^{-2})$ 
& $O(m_\pi^{-2} \De^{-1})$ & $O(\De^{-3})$  &  &  Ref.~\cite{Hol00} 
& Ref.~\cite{BGLMN}\\
\hline
$\alpha_{E\nu}$ & 2.2 & 0 & $-5.5$  &$-1.5$ &0.7& $ -4.1 $ 
&$-4.0$ &$-3.8$  \\
$\be_{M\nu}$ & 3.5 & 5.0  & $-1.1$  & 1.4 & $-0.2$& 8.5  & 9.3&9.1\\
$\al_{E2}$ & 20.8 & 0& 0& 6.7 & $-0.8$& 26.7 & 29.1 & 27.5  \\
$\be_{M2}$ & $-8.9$& 0& $-2.0$  & $-5.5$ & 0.5& $-15.9$  & $-24.1$ & $-22.4$\\
\hline\hline
\end{tabular}
\caption{Pieces of fourth-order spin-independent nucleon
polarizabilities which scale with negative 
powers of $m_\pi$ and $\De$ together with a comparison to
dispersion-relation analyses. 
All values are in units of $10^{-4}$ fm$^5$.}
\tablab{pols4}
\end{table}

\begin{table}[htb]
\begin{tabular}{||c|r|r||r|r||}
\hline
&  \multicolumn{2}{|c||}{$\De$ excitation} & 
\multicolumn{2}{|c||}{Sum} \\
\cline{2-5} 
  & LO & NLO  &  Proton & Neutron \\
\hline
$\,\ga_1\,$ &0 & $3.1$ &$-18.4$ &28.3\\
$\,\ga_2\,$ &$-2.5$ & $0$ &$-1.3$ &$-0.8$\\
$\,\ga_3\,$ &0 & $-1.5$ &$10.1$&  $-12.2$\\
$\,\ga_4\,$ &2.5 & $1.5$ &$-6.7$& $15.4$\\
\hline
$\,\ga_0 \,$ & $-2.5$ &0.1 &$-3.7$ & $-1.7$ \\
$\,\ga_\pi\,$ &   $2.5$ & 6.2 &  $-33.0$ &$58.4$ \\
\hline
$\alpha_{E\nu}$ & 0& $-3.3$ & $-1.8$ & $-1.8$\\
$\be_{M\nu}$ &6.6&$-1.1$& 10.7 &10.7\\
$\al_{E2}$ & 0 & 0 & 26.7 & 26.7\\
$\be_{M2}$ & 0 &$-2.7$&$-16.6$ & $-16.6$\\
\hline\hline
\end{tabular}
\caption{
Results for $\De$-excitation pieces of the
spin and fourth-order polarizabilities using an alternative
set of $\ga N\De$ parameters: $g_M=3$, $g_E=-1$. The last two columns
indicate how the total prediction is changed.}
\tablab{altpols}
\end{table}

\end{document}